\documentstyle[preprint,aps,prd,eqsecnum]{revtex}
\tighten

\def\bfl{\begin{flushleft}}
\def\efl{\end{flushleft}}
\def\bfr{\begin{flushright}}
\def\efr{\end{flushright}}
\def\bc{\begin{center}}
\def\ec{\end{center}}
\def\be{\begin{equation}}
\def\ee{\end{equation}}
\def\ba{\begin{eqnarray}}
\def\ea{\end{eqnarray}}
\def\nn{\nonumber }
\def\lb#1{\label{#1}}

\def\text#1{\mbox{#1}}
\def\drm{d}
\def\Adequa{\Longleftrightarrow}
\def\schrod{Schr\"oedinger  }
\def\vphi{\varphi}
\def\phik#1#2{\varphi_{#1 sol}^{#2}}
\def\Mass{{\cal M}}

\def\Sign#1{\, \text{sign}\left(#1\right) }
\def\Der#1#2{\,\frac{\partial #1}{\partial #2}}

\def\Int#1#2{\, \int\limits_{#1}^{#2}}

\def\Arcsh#1{\,\text{arcsinh} \left[#1\right]}
\def\Cosh#1#2{\, \text{cosh}^{#1}\left(#2 \right) }
\def\Sech#1#2{\, \text{sech}^{#1}\left(#2 \right) }
\def\HypGF#1#2#3#4{\,\text{F}\left(#1,\,#2,\,#3;\,#4 \right) } 

\begin{document}

\draft

\title{
~~~~~~~~~~~~~~~~~~~~~~
{\small J. Phys. G: Nucl. Part. Phys. 25 (1999) 2177-2187}\\
~\\
Zero-brane approach to study of particle-like solitons
in classical and quantum Liouville field theory
      }

\author{Konstantin G. Zloshchastiev}
\address{~\\E-mail: zlosh@email.com.\\
URL(s): http://zloshchastiev.webjump.com, http://zloshchastiev.cjb.net\\~\\}

\date{Received April 3, 1999}
\maketitle

\begin{abstract}
The effective p-brane action
approach is generalized for arbitrary scalar field
and applied for the Liouville theory near a particle-like
solution.
It was established that this theory has the remarkable features
discriminating it from the theories studied earlier.
Removing zero modes we obtain the effective action describing the solution
as a point particle with curvature,
quantize it as the theory 
with higher derivatives and calculate the quantum corrections to  
mass.
\end{abstract}


\narrowtext

\section{Introduction}\lb{s-i}

Appeared more than a century ago \cite{lio}
the Liouville model have a wide application both
in conventional physics \cite{dpp} and in
such new-fashioned regions of modern physics as 
Chern-Simons theory and gauge gravity \cite{jac}, 
string and conformal field theory \cite{kpz}, 
Wess-Zumino-Witten model \cite{os}, 
superstring and M-theory \cite{lpx},
Seiberg-Witten approach \cite{ns}, to mention just few 
examples and references.

The aim of this paper is to study the Liouville theory
in the neighborhood of the particle-like solution
within the frameworks of the brane approach, which 
consists in the 
constructing of the effective action where the non-minimal
terms (first of all, depending on the world-volume curvature) are 
induced by the field fluctuations.
Then the required action evidently arises after nonlinear 
reparametrization
of an initial theory when excluding zero field oscillations.

The paper is arranged as follows.
In Sec. \ref{s-ps} we study the Liouville particle-like soliton
on the classical level.
In Sec. \ref{s-ea} we generalize and enhance the approach developed by 
Kapustnikov and Pashnev \cite{kpp}, for 
arbitrary scalar fields and apply it for the model in question.
Sec. \ref{s-q} is devoted to quantization of the obtained
zero-brane action as a constrained theory with higher derivatives.
In result we obtain the \schrod wave equation effectively describing
wave function and mass spectrum of a point particle with curvature,
calculate the null and first excited levels to derive the
quantum corrections to the mass of the studied soliton.
Conclusions are made in Sec. \ref{s-c}.

\section{Particle-like solutions}\lb{s-ps}

We start with the action  
\begin{equation}
S[\varphi] = \frac{1}{2} \int \drm^2 x 
\left[
(\partial_m \varphi)^2 +
m \text{e}^{2\beta\varphi} + \zeta
\right],                                                     \lb{eq2.1} 
\end{equation}
where $m$, $\beta$ and $\zeta$ are constants. 
The equation of motion in the class of solitary waves is
\ba                                                           
\varphi_{\rho\rho} + 
m\beta\, \text{e}^{2\beta\varphi} = 0,                        \lb{eq2.2}
\ea
$\rho=\gamma(x-v t)$, $\gamma= 1/\sqrt{1-v^2}$.
Its general solution has the form
\be                                                           \lb{eq2.3}
\phik{}{}(\rho) = \frac{1}{\beta}
\Arcsh{
       \frac{C - \Cosh{2}{\sqrt{m C}\beta\rho}}
            {2\sqrt{C} \Cosh{}{\sqrt{m C}\beta\rho}}
      },
\ee
where the $C\not=0$ is another integration constant, 
all the constants are assumed to be 
such that the potential is real everywhere
on the axis.
The corresponding energy density,
\[
\varepsilon_{sol} (x,t) = \frac{m}{2}
\left[
C - \zeta/m - 2 C \Sech{2}{\sqrt{m C}\beta\rho}
\right],
\]
turns to be localized if we suppose
\be                                                           \lb{eq2.4}
C = \zeta/m.
\ee
Then the total energy is finite despite the potential diverges
at the infinity, and can be rewritten in the explicit
relativistic point-particle form
\be                                                           \lb{eq2.5}
E_{\text{class}} = 
\Int{-\infty}{\infty} \varepsilon (x,t)\, \drm x = \gamma\mu,\ \
\mu = \frac{2\sqrt{\zeta}}{\beta},
\ee
hence it is interesting to note
that the additive constant $\zeta$ (in spite of it 
does not appear in the equation of motion) affects on the most important
properties of the particle-like solitons (\ref{eq2.3}) by virtue
of the localization requirement (\ref{eq2.4}).
This is the key difference of the Liouville theory from those
studied earlier, viz., $\varphi^4$ \cite{kpp}, 
$\varphi^3$ \cite{zlo006} and sine-Gordon.  
Among the other features one can mention the initial  
$\varphi \to -\varphi$ symmetry breaking 
(like $\varphi^3$-model) and
that the theory does not admit the trivial vacuum state $\varphi=0$.

\section{Effective action}\lb{s-ea}

In this section we take into account the field fluctuations
in the neighborhood of the solution (\ref{eq2.3}), (\ref{eq2.4}) 
and construct the nonlinear effective zero-brane (non-minimal
point particle) action.
However, firstly, perfecting and generalizing the method \cite{kpp},
we develop some general approach to such theories. 

\subsection{General formalism}\lb{s-ea-gf}

Let us consider the action describing arbitrary
one-scalar field
\begin{equation}
S[\vphi] = \int  L(\vphi)\, \drm^2 x,                        \lb{eq3.1} 
\end{equation}
\be                                                          \lb{eq3.2}
L (\vphi) = \frac{1}{2} 
 (\partial_n \varphi) (\partial^n \varphi) -
U (\vphi).
\ee
The corresponding equation of motion is
\be                                                          \lb{eq3.3}
\partial^n \partial_n \varphi + U_1(\vphi) = 0,
\ee
where we defined                                   
\[
U_1(\vphi) = \Der{U(\vphi)}{\varphi},~~
U_{2}(\vphi) = \Der{^2\, U(\vphi)}{\varphi^2 }.
\]
Suppose, we have a solution in the class of solitary waves
\be                                                           \lb{eq3.4}
\phik{}{}(\rho) = \phik{}{}
\left( \gamma ( x-v t)
\right),  
\ee
having the localized energy density
\be                                                           \lb{eq3.5}
\varepsilon (\vphi) =  
\Der{L(\vphi)}{(\partial_0\vphi)} \partial_0\vphi - L(\vphi),
\ee
and finite mass integral
\be                                                            \lb{eq3.6}
\mu = \int\limits_{-\infty}^{+\infty}
\varepsilon (\phik{}{})\ \drm \rho =
-\int\limits_{-\infty}^{+\infty}
L (\phik{}{})\ \drm \rho < \infty,
\ee
coinciding with the total energy up to the Lorentz factor $\gamma$.

Let us change to the set of the collective coordinates 
$\{\sigma_0=s,\ \sigma_1=\rho\}$ such that
\be                                                      
x^n = x^n(s) + e^n_{(1)}(s) \rho,\ \              
\varphi(x,t) = \widetilde \varphi (\sigma),     
\ee
where $x^n(s)$ turn to be the coordinates of a (1+1)-dimensional point
particle, $e^n_{(1)}(s)$ is the unit spacelike vector orthogonal
to the world line.
Hence, the action (\ref{eq3.1}) can be rewritten in new coordinates as
\be                                                         \lb{eq3.8}
S[\widetilde \varphi] = 
\int L (\widetilde \varphi) \,\Delta \ \drm^2 \sigma,
\ee
\[
L (\widetilde \varphi) = \frac{1}{2} 
\left[
      \frac{(\partial_s \widetilde\varphi)^2}{\Delta^2} - 
                  (\partial_\rho \widetilde\varphi)^2
\right]
- U (\widetilde \varphi),
\]
where
\[
\Delta = \text{det} 
\left|
\left|
      \Der{x^n}{\sigma^l}
\right|
\right|
= \sqrt{\dot x^2} (1- \rho k),
\]
and $k$ is the curvature of a particle world line
\be                                                            \lb{eq3.9}
k = \frac{\varepsilon_{l n} \dot x^l \ddot x^n}{(\sqrt{\dot x^2})^3},
\ee
where $\varepsilon_{l n}$ is the unit antisymmetric tensor.
This new action contains the redundant degree of freedom which 
eventually
leads to appearance of the so-called ``zero modes''.
To eliminate it we must constrain the model
by means of the condition of vanishing of the functional derivative with
respect to field fluctuations about a chosen static solution,
and in result we will obtain the required effective action.

So, the fluctuations of the field $\widetilde\varphi (\sigma)$ in the 
neighborhood of the static solution $\phik{}{} (\rho)$
are given by the expression
\be                                               
\widetilde\varphi (\sigma) = 
\phik{}{} (\rho) + \delta \varphi (\sigma).
\ee
Substituting them into eq. (\ref{eq3.8}) and considering the static
equation of motion (\ref{eq3.3}) for $\phik{}{}(\rho)$ we have
\ba                                          \lb{eq3.11}
S[\delta \vphi] 
&=& \int d^2 \sigma \ 
   \Biggl\{\Delta 
        \Biggl[L( \phik{}{}) +
               \frac{1}{2} 
               \Biggl( 
                  \frac{\left(\partial_s \ \delta \varphi \right)^2}
                       {\Delta^2} 
                  -  
                  \Bigl( 
                        \partial_{\rho}  \delta \varphi 
                  \Bigr)^2 - \nn\\
&&                  U_{2} ( \phik{}{}) 
                  \delta \varphi^2
               \Biggr) 
        \Biggr]
        - k \sqrt{\dot x^2} \phik{}{\prime} \delta \varphi
        + O (\delta \varphi^3)                            
    \Biggr\} + [\rho\text{-surface terms}],                                            
\ea
\[
L( \phik{}{}) = 
 -\frac{1}{2}  \phik{}{\prime\, 2} - U ( \phik{}{}),
\]
where prime means the derivative with respect to $\rho$.
Extremalizing this action with respect to 
$\delta \varphi$ one can obtain 
the system of equations in partial derivatives for field fluctuations:
\be
\left(
     \partial_s \Delta^{-1} \partial_s -
     \partial_{\rho} \Delta \partial_{\rho} 
\right) \delta\varphi
+\Delta  U_{2} ( \phik{}{}) \delta\varphi
+ \phik{}{\prime} k\sqrt{\dot{x}^2} =
O(\delta \varphi^2),
\ee
which has to be the constraint removing a redundant degree of
freedom.
Supposing $\delta\varphi (s,\rho) = k(s) f(\rho)$, in the 
linear approximations
$\rho k\ll 1$ (which naturally guarantees also
the smoothness of a world line at $\rho \to 0$) 
and $O(\delta\varphi^2)=0$ we obtain a system
of the ordinary derivative equations
\ba  
&&\frac{1}{\sqrt{\dot{x}^2}} \frac{d}{ds} 
\frac{1}{\sqrt{\dot{x}^2}} \frac{dk}{ds} +ck = 0,          \lb{eq3.13}\\
&&-f'' + 
\left( 
      U_{2} ( \phik{}{}) - c 
\right) f + \phik{}{\prime} = 0,                        \lb{eq3.14}
\ea
where $c$ is the constant of separation.
Searching for a solution of the last equation in the form
\be                                                       \lb{eq3.15}
f = g + \frac{1}{c} \phik{}{\prime},
\ee
we obtain the homogeneous equation
\be  
-g'' + 
\left( 
      \frac{\phik{}{\prime\prime\prime}}
           {\phik{}{\prime}}
      - c 
\right) g = 0.                                           \lb{eq3.16}
\ee
Strictly speaking, the explicit form of $g (\rho)$ is not significant 
for us, because we always can suppose integration constants to be zero
thus restricting ourselves by the special solution.
Nevertheless, the homogeneous equation should be considered as the 
eigenvalue problem for $c$ (see below).

Substituting the found function $\delta\varphi = k f$ back 
in the action 
(\ref{eq3.11}), we can rewrite it in the explicit p-brane form
\be                                \lb{eq3.17}
S_{\text{eff}} = 
S_{\text{eff}}^{\text{(class)}} + S_{\text{eff}}^{\text{(fluct)}} =
- \int \drm s \sqrt{\dot x^2} 
\left(
       \mu + \alpha k^2
\right),
\ee
describing a point particle with curvature,
where $\mu$ was defined in (\ref{eq3.6}), and
\be                                               \lb{eq3.18}
\alpha = 
\frac{1}{2}  \int\limits_{-\infty}^{\infty} 
f \phik{}{\prime} \ \drm \rho
+
\frac{1}{2}  
\int\limits_{-\infty}^{+\infty} 
\left(
f f^{\prime}
\right)^\prime \ \drm \rho.
\ee
Further, from the static equation (\ref{eq3.3}) we
obtain the expression
\be                                                \lb{eq3.19}
\left(
\phik{}{\prime\prime} -
U_1(\phik{}{})
\right) 
\phik{}{\prime} = 0,
\ee
which can be rewritten as 
\be                                                \lb{eq3.20}
\phik{}{\prime 2} 
= 2 U(\phik{}{}(\rho)).
\ee
Considering eqs. (\ref{eq3.5}), (\ref{eq3.6}), (\ref{eq3.15}), 
(\ref{eq3.19}) and (\ref{eq3.20}), the expression for $\alpha$ can be 
written in the simple form
\be
\alpha = \frac{\mu}{2 c} + \frac{1}{2 c^2}
\int\limits_{-\infty}^{+\infty} 
U^{\prime\prime}(\phik{}{}(\rho)) \ \drm \rho,
\ee
where the second term can be integrated as a full derivative,
and vanishes when $|\phik{}{\prime}(\rho)| \leq O(1)$ at infinity.
Even it does not happen,
we always can include this term into 
the surface terms of the action (\ref{eq3.11}).
Thus, we obtain the final form of the effective p-brane action of
the theory
\be                                \lb{eq3.22}
S_{\text{eff}} = 
- \mu \int \drm s \sqrt{\dot x^2} 
\left(
       1 + \frac{1}{2 c} k^2
\right).
\ee
It is straightforward to derive the corresponding 
equation of motion in the Frenet basis
\be
\frac{1}{\sqrt{\dot x^2}}
\frac{\drm}{\drm s}
\frac{1}{\sqrt{\dot x^2}}
\frac{\drm k}{\drm s} +
\left(c - \frac{1}{2} k^2
\right) k = 0,
\ee
hence one can see that eq. (\ref{eq3.13}) was nothing
but this equation in the linear approximation $k \ll 1$,
as was expected.

Thus, the only problem which yet demands on the 
resolving is the determination of eigenvalue $c$.
It turns to be the Stourm-Liouville problem for eq. (\ref{eq3.16}) 
provided some chosen boundary conditions.
If one supposes, for instance, the finiteness of $g$ at infinity
then the $c$ spectrum turns to be discrete.
Moreover, it often happens that $c$ has only one or two admissible
values \cite{kpp,zlo006}.
In any case we are needed in the exact value of $c$ hence for
each concrete case
eq. (\ref{eq3.16}) should be resolved as exactly as possible.

\subsection{Application for Liouville model}\lb{s-ea-af}

We will suppose the final p-brane
action (\ref{eq3.22}) and will determine its parameters.
We already have $\mu$ derived in (\ref{eq2.5}), and 
the eigenvalue $c$ remains to be 
the only unknown parameter.
As a boundary condition for the 
Stourm-Liouville problem (\ref{eq3.16}),
we require
\be                                                          \lb{eq3.24}
g(+\infty) - g(-\infty) = O (1),
\ee
whereas (\ref{eq3.16}) reads
\be                                                          \lb{eq3.25}
g'' + 
\left(
c + \frac{\zeta\beta^2}{\Cosh{2}{\sqrt{\zeta}\beta\rho}}
\right)g = 0.
\ee
According to the proven theorem (see Appendix \ref{a-a}), 
the only admissible non-zero $c$ is
\be
c=-\zeta\beta^2,
\ee
hence the effective zero-brane action of the Liouville model near
particle-like solution (\ref{eq2.3}), (\ref{eq2.4}) 
with fluctuational corrections is
\be                                                   \lb{eq3.27}
S_{\text{eff}} = 
- \mu \int \drm s \sqrt{\dot x^2} 
\left(
       1 - \frac{k^2}{2 \zeta\beta^2} 
\right).
\ee
In the next section we will quantize it to obtain the quantum 
corrections to the mass of the solution.

\section{Quantization}\lb{s-q}

In the previous section we obtained a classical 
effective action for the model in question.
Thus, to quantize it we must consecutively construct the 
Hamiltonian structure of dynamics of the point particle with 
curvature \cite{ner}.
From the p-brane action (\ref{eq3.22}) 
and definition of the world-line curvature 
one can see that we have the 
theory with higher derivatives \cite{dhot}.
Hence, below we will treat the coordinates and momenta as the 
canonically independent coordinates of phase space.
Besides, the Hessian matrix constructed 
from the derivatives with respect to accelerations,
\[
M_{a b} = 
\left|
\left|
\Der{^2 L_{\text{eff}}}{\ddot x^a \partial \ddot x^b}
\right|
\right|,
\] 
appears to be singular that says about the presence of the
constraints on phase variables of the theory.

As was mentioned, the phase space consists of the two pairs of 
canonical variables:
\ba
&&x_m,\ \ p_m = \Der{L_{\text{eff}}}{q^m} - \dot \Pi_m, \\
&&q_m = \dot x_m,\ \ \Pi_m =\Der{L_{\text{eff}}}{\dot q^m},
\ea
hence we have
\ba
&&p^n = - e^n_{(0)} \mu 
\left[
      1-  \frac{1}{2 c}
\right] +
\frac{\mu}{c}
\frac{e^n_{(1)} }{\sqrt{q^2}} \dot k,  \\
&&
\Pi^n = - 
\frac{\mu}{c}
\frac{e^n_{(1)}}{\sqrt{q^2}} k,
\ea
where the components of the Frenet basis are
\[
e^m_{(0)} = \frac{\dot x^m}{\sqrt{\dot x^2}},\
e^m_{(1)} = - \frac{1}{\sqrt{\dot x^2}} \frac{\dot e^m_{(0)}}{k}.
\]
There exist the two primary constraints of first kind
\ba
&&\Phi_1 = \Pi^m q_m \approx 0, \\
&&\Phi_2 = p^m q_m + \sqrt{q^2} 
\left[
      \mu + \frac{c}{2 \mu} q^2 \Pi^2
\right]  \approx 0,
\ea
besides we should add the proper time gauge condition,
\be
G = \sqrt{q^2} - 1 \approx 0,
\ee
to remove the non-physical gauge degree of freedom.
Then, when introducing the new variables,
\be
\rho = \sqrt{q^2},\ \ v = 
\text{arctanh} 
\left(
      p_{(1)}/p_{(0)}
\right),
\ee
the constraints can be rewritten in the form
\ba
&&\Phi_1 = \rho \Pi_\rho, \nn\\
&&\Phi_2 = \rho 
\left[
      -\sqrt{p^2} \cosh{v} + \mu -
      \frac{c}{2 \mu}
      \left( 
            \Pi^2_v - \rho^2 \Pi^2_\rho
      \right)
\right],                                       \\
&&G=\rho-1, \nn
\ea
hence finally we obtain the constraint
\be
\Phi_2 = 
      -\sqrt{p^2} \cosh{v} + \mu -
      \frac{c}{2 \mu}
      \Pi^2_v \approx 0,                                       
\ee
which in the quantum theory ($\Pi_v = - i \partial/\partial v$) 
yields 
\[
\widehat\Phi_2 |\Psi\rangle =0.
\]
As was shown in Ref. \cite{kpp}, the constraint $\Phi_2$ on the 
quantum level admits several coordinate representations that,
generally speaking, lead to different 
nonequivalent theories, therefore,
the choice between the different forms of 
$\widehat\Phi_2$ should be based on the physical relevance.
Then the physically admissible
equation determining quantum dynamics of the quantum
kink and bell particles has the form:
\be                                                            \lb{eq4.11}
[ \widehat H-\varepsilon] \Psi(\zeta) = 0, 
\ee 
\be
\widehat H =  -\frac{\drm^2}{\drm \zeta^2} +
  \frac{B^2}{4}
  \sinh{\! ^2 \zeta}
  -B
  \left(
        S+\frac{1}{2}
  \right)
  \cosh{\zeta},                                             
\ee
where
\ba
&&\zeta=v/2,\ \sqrt{p^2} = \Mass, \nn\\
&&B= 8 \sqrt{
            \frac{\mu \Mass}{c}  
            },                                              \lb{eq4.13}\\
&& \varepsilon = \frac{8 \mu^2}{c}  
\left(
      1 - \frac{\Mass}{\mu}
\right), \nn
\ea
and $S=0$ in our case.

As was established in the works \cite{raz,zu}, 
SU(2) has to be the dynamical symmetry
group for this Hamiltonian which can be rewritten in the form of
the spin Hamiltonian
\be
\widehat H= -S^2_z - B S_x,                                 
\ee
where the spin operators,
\ba
&&S_x = S \cosh{\zeta} - \frac{B}{2} \sinh{\!^2 \zeta} - \sinh{\zeta} 
\frac{\drm}{\drm\zeta},  \nn \\
&&S_y = i         \left\{
               -S \sinh{\zeta} + \frac{B}{2} \sinh{\zeta}\cosh{\zeta} + 
\cosh{\zeta} \frac{\drm}{\drm\zeta} \right\},   \\
&&S_z =          \frac{B}{2} \sinh{\zeta}
        + \frac{\drm}{\drm\zeta},              \nn
\ea
satisfy with the commutation relations
\[
[S_i,~S_j] = i \epsilon_{ijk} S_k,                       
\]
besides
\[
S_x^2+S_y^2+S_z^2 \equiv S (S+1).                         
\]
In this connection it should be noted that though the reformulation of 
some interaction 
concerning the coordinate degrees of freedom in terms of
spin variables is widely used (e.g., in the theories with the
Heisenberg Hamiltonian, see Ref. \cite{lp}), it has to be just
the physical approximation as a rule,
whereas in our case the spin-coordinate correspondence is exact.

Further, 
at $S\geq 0$ there exists an irreducible ($2 S+1$)-dimensional 
subspace of the representation space of the su(2) Lie algebra, which is
invariant with respect to these operators.
Determining eigenvalues and eigenvectors of the spin Hamiltonian
in the matrix representation 
which is realized in this subspace, one can prove 
that the solution of eq. (\ref{eq4.11}) is the function
\ba
\Psi (\zeta) =\exp{
                   \left(
                         -\frac{B}{2} \cosh{\zeta}
                   \right)
                  }
              \sum_{\sigma=-S}^{S}
              \frac{c_\sigma}
                   {
                    \sqrt{
                          (S-\sigma)\verb|!|~
                          (S+\sigma)\verb|!|
                         }
                   }
              \exp{
                   \left(
                         \sigma \zeta
                   \right)
                  }, 
\ea
where the coefficients $c_\sigma$ are the solutions of 
the system of linear equations
\[
\biggl(
       \varepsilon+\sigma^2
\biggr)c_\sigma + \frac{B}{2}
\biggl[
       \sqrt{(S-\sigma)(S+\sigma+1)}~ c_{\sigma+1}             
+ \sqrt{(S+\sigma)(S-
\sigma+1)}~ c_{\sigma-1}
\biggr] = 0,
\]
\[
c_{S+1} = c_{-S-1}=0,~~\sigma=-S,~-S+1,...,~S.            
\]
However, it should be noted that these expressions give only the 
finite number of exact solutions which is equal to the dimensionality of
the invariant subspace 
(this is the so-called QES, quasi-exactly solvable, system).
Therefore, for the spin $S=0$ we can find only the ground state wave
function and eigenvalue:
\be
\Psi_0 (\zeta) = C_1 
\exp{
    \left(
           - \frac{B}{2} \cosh{\zeta}
    \right)
    },\ 
\varepsilon_0 = 0.
\ee
Hence, we obtain that the ground-state mass of 
the quantum particle with curvature coincides with the classical one,
\be                                                         \lb{eq4.18}
\Mass_0 = \mu,
\ee
as was expected (strictly speaking, it coincides up to the sign which
is insufficient, see below).

Further, in absence of exact wave functions for more excited 
levels one can find
the first (small) quantum correction to mass 
in the approximation of the quantum harmonic oscillator.
It is easy to see that at $B \geq 1$
the (effective) potential 
\be
V(\zeta) = 
\left(
      \frac{B}{2}
\right)^2 \text{sinh}^2 \zeta
-
\frac{B}{2} \cosh{\zeta}
\ee
has the single minimum
\[
V_{\text{min}} = - B/2 \ \ \text{at} \ \ \zeta_{\text{min}}=0.
\]

Then following to the $\hbar$-expansion technique we shift the origin of 
coordinates in the point of minimum (to satisfy 
$\varepsilon = \varepsilon_0 = 0$ in absence
of quantum oscillations), and expand $V$ in the 
Taylor series to second order
near the origin thus reducing the model to the oscillator
of the unit mass, energy $\varepsilon/2$ and oscillation frequency
\[
\omega = \frac{1}{2} \sqrt{B(B-1)}.
\]
Therefore, the quantization rules yield the discrete spectrum
\be
\varepsilon =  \sqrt{B(B-1)} (n + 1/2) 
+ O (\hbar^2),\ \ n=0,\ 1,\ 2, ...,
\ee
and the first quantum correction to particle masses will be
determined by the lower energy of oscillations:
\be                                                        
\varepsilon = \frac{1}{2} \sqrt{B(B-1)} + O (\hbar^2),
\ee
that gives the algebraic equation for $\Mass$ as a function of $m$
and $\mu$.

We can easily resolve it in the approximation 
\be                                                        \lb{eq4.22}      
B \gg 1 \ \Adequa \ c/\mu^2 \to 0,
\ee
which is admissible for the major physical cases, and obtain
\be                                                         
\varepsilon = \frac{B}{2} + O (\hbar^2 c/\mu^2),
\ee
that after considering of eqs. (\ref{eq4.13}) and (\ref{eq4.18}) yields
\be
(\Mass-\mu)^2 = \frac{c \Mass}{4 \mu} + O (\hbar^2 c/\mu^2).
\ee
Then one can seek for mass in the form $\Mass=\mu+\delta$ 
($\delta \ll \mu$), and 
finally we obtain the mass of a particle with curvature (\ref{eq3.22})
with first-order quantum corrections (considering
that $\Mass$ is always defined up to a sign, 
besides $\mu$ is defined up to a sign as well)
\[
\Sign{\Mass} \Mass = \mu \pm \frac{\sqrt{\Sign{\Mass}c}}{2} 
+ O(\hbar^2 c/\mu^2),
\]
where $\Sign{\Mass}$ is chosen such that after all we have
a positive value
\be                                             \lb{eq4.25}
\Mass = |\mu| 
\pm \frac{\sqrt{|c|}}{2} + O(\hbar^2 c/\mu^2),
\ee
i. e., quantization procedure restores the positivity of mass.

The nature of the justified choice of the root sign before the second term
is not so clear as it seems for a first look,
because there exist the two interfering points of view.
The first (physical) one is: 
if we apply this formalism for the one-scalar $\varphi^4$
model \cite{kpp} and compare the result with that obtained in other
ways \cite{raj}, we should suppose the sign ``$+$'' (or, at least, 
Rajaraman made no mentions upon the choice of signs).
However, the second, mathematical, counterargument is as follows: 
the known exact spectra of the operators with the QES potentials 
like (\ref{eq4.11}) are
split, as a rule by virtue of radicals, hence the signs ``$\pm$''
might approximately represent such a bifurcation and thus
should be unharmed.
If it is really so, quantum fluctuations should divide the 
classically unified particle with
curvature into several subtypes with respect to mass.

Let us apply these results for our case.
One can see that $\mu$ is indeed defined up to a sign.
Therefore, considering eqs. (\ref{eq2.5}), (\ref{eq3.27}) 
and (\ref{eq4.25}), the mass of the quantum 
particle-like solution (\ref{eq2.3}) in the first approximation is
\be
\Mass = 
\sqrt{\zeta}
\left(
      \frac{2}{\beta} 
      \pm
      \frac{\beta}{2} 
\right),
\ee
where for beauty we have omitted the magnitude symbol but
suppose the r.h.s. to be positive.
The problem of the obtaining of further corrections appears to be
the mathematically standard
Stourm-Liouville problem for the Razavi potential, all 
the more so it is well-like on the whole axis and hence admits
only the bound states with a discrete spectrum.

Finally we note that because of the Liouville model does not contain
the vacuum state $\varphi=0$, the obtained spectrum is 
nonperturbative and can not be derived by virtue of the standard
perturbation theory starting from the vacuum sector.

\section{Conclusion}\lb{s-c}

Let us enumerate the main results obtained.
It was shown that the Liouville field theory admits
the particle-like field solution which (weakly) diverges
at infinity but nevertheless have the localized energy.
Besides, its mass depends on the constant which is
additive to the Lagrangian and, therefore, does not appear
in the equation of motion.

Further, considering field fluctuations in the neighborhood of 
this soliton we ruled out the action for it as a non-minimal 
point particle with curvature,
thereby we have generalized and polished the procedure of
obtaining of brane actions.
When quantizing this action as the constrained 
theory with higher derivatives, 
it was shown that the resulting \schrod equation is the
special case of the Razavi equation
having SU(2) dynamical symmetry group in the ground state.
Finally, we found the first quantum correction to 
mass of the solution in question which could not be 
calculated by means of series of the perturbation theory.

\appendix
\section{Eigenvalue theorem}\lb{a-a}

{\it Theorem.}
The bound-state singular Stourm-Liouville problem
\be                                               \lb{eq-a1}
-f''(u) +  
\left(
      1 - 2\, \text{sech}^2 u
\right)f(u) - c f(u)  = 0,
\ee
\be                                               \lb{eq-a2}
f(+\infty) = f(-\infty) = O(1),
\ee 
has only the two sets of eigenfunctions and eigenvalues
\ba
&&f_0 = K_0\, \text{sech}\, u,\ c_0=0,           \nn\\
&&f_1 = K_1 \tanh{u},\ c_1=1.                \nn
\ea
where $K_i$ are arbitrary integration constants.

{\it Proof.}
Performing the change $z =  \text{cosh}^2 u$, we rewrite the 
conditions of the theorem in the form
\be                                                \lb{eq-a3}
2 z (z-1) f_{z z} + (2 z -1) f_{z} -
\left(
      \frac{\widetilde c}{2} - \frac{1}{z}
\right) f = 0,
\ee
\be                                               \lb{eq-a4}
f(1) = 0,\ \ f(+\infty) = O(1),
\ee 
where $\widetilde c = 1 - c$.
The general integral of eq. (\ref{eq-a3}) 
can be expressed in terms of the hypergeometric functions
\[
f = 
\frac{C_1}{\sqrt{z}}
\HypGF{\frac{-1-\sqrt{\widetilde c}}{2}}
      {\frac{-1+\sqrt{\widetilde c}}{2}}
      {-\frac{1}{2}}{z}
+ C_2 z
\HypGF{1-\frac{\sqrt{\widetilde c}}{2}}
      {\frac{1+\sqrt{\widetilde c}}{2}}
      {\frac{5}{2}}{z}.
\]
Using the asymptotics of the hypergeometric functions in the neighborhood
$z=1$, it is straightforward to derive that the first from the conditions
(\ref{eq-a4}) will be satisfied if we suppose
\be                                                         \lb{eq-a5}
\frac{1}{C_1} f^{(\text{reg})} = 
\frac{1}{\sqrt{z}}
\HypGF{-1-\frac{\sqrt{\widetilde c}}{2}}
      {-1+\frac{\sqrt{\widetilde c}}{2}}
      {-\frac{3}{2}}{z}
- C^{(\text{reg})} z
\HypGF{\frac{3-\sqrt{\widetilde c}}{2}}
      {\frac{3+\sqrt{\widetilde c}}{2}}
      {\frac{7}{2}}{z},
\ee
where
\[
C^{(\text{reg})} = \sqrt{\widetilde c}  
(\widetilde c - 1)
\tan{
\left(
    \frac{\pi\sqrt{\widetilde c}}{2}
\right)
    }.
\]
Further, to specify the parameters at which this function satisfies
with the second condition (\ref{eq-a4}) we should consider the 
asymptotical behavior of $f^{(\text{reg})}$ near infinity.
We have
\be                  
\frac{1}{C_1} f^{(\text{reg})} (z \to \infty) =
\frac{2\, \breve\gamma}{\pi^{3/2}}
(-1)^{1+\sqrt{\widetilde c}/2} 
\tan{
     \left( 
           \frac{\pi \sqrt{\widetilde c}}{2}
     \right)
    }
\sin{
     \left( 
           \frac{\pi \sqrt{\widetilde c}}{2}
     \right)
    }
z^{\sqrt{\widetilde c}/2}
\biggl[ 
       1+ O(1/z)
\biggr],
\ee
where
\[
\breve\gamma = \Gamma (\sqrt{\widetilde c})
\left[
      i \sqrt{\widetilde c} (\widetilde c - 1) 
      \Gamma (-1/2 - \sqrt{\widetilde c}/2)
      \Gamma (\sqrt{\widetilde c}/2)
       - 8\,
      \Gamma (1 - \sqrt{\widetilde c}/2)
      \Gamma (3/2 - \sqrt{\widetilde c}/2)
\right].
\]
From this expression it can easily be seen that 
$ f^{(\text{reg})}$ diverges at infinity everywhere except
perhaps the points:
\[
\widetilde c = (2 n)^2 = 0,\ 4,\ 16, ...,\ \ 
\text{and} \ \ \widetilde c = 1,
\]
which demand on an individual consideration.
From eq. (\ref{eq-a3}) we have 
\[
f_{\widetilde c = 0} = 
C_1 
\sqrt{1-\frac{1}{z}} 
+ 
C_2 
\left[
      i - 
      \sqrt{1-\frac{1}{z}} 
      \arcsin{\sqrt{z}}
\right],
\]
\[
f_{\widetilde c = 1} = 
\frac{C_1}{\sqrt{z}} 
+ 
C_2 
\left[
      \sqrt{1-\frac{1}{z}} -
      i \frac{\arcsin{\sqrt{z}}}{\sqrt{z}}
\right],
\]
\[
\widetilde f_{\widetilde c = 4} = 
C_1 
\sqrt{1-\frac{1}{z}} 
\left(
      2 z + 1
\right)
+ C_2 z,
\]
\[
f_{\widetilde c = 16} = 
C_1 
\sqrt{1-\frac{1}{z}} 
\left(
      24 z^2 - 8 z - 1
\right)
+ C_2 z (1-6 z/5),
\]
and so on.
By induction it is clear that at 
$\widetilde c \geq 4$
there are no $C_i$ at which $ f$ would satisfy with the requirements
(\ref{eq-a2}).

\def\CJP{Czech. J. Phys.}
\def\CMPh{Commun. Math. Phys.}
\def\CQG {Class. Quant. Grav.}
\def\DANt {Sov. Phys. Dokl.}
\def\JMP{J. Math. Phys.}
\def\JPh{J. Phys.}
\def\FP{Fortschr. Phys.}
\def\GRG {Gen. Relativ. Gravit.}
\def\LMPh {Lett. Math. Phys.}
\def\MPL {Mod. Phys. Lett.}
\def\NPh  {Nucl. Phys.}
\def\PhE  {Phys.Essays}
\def\PhL  {Phys. Lett.}
\def\PhR  {Phys. Rev.}
\def\PhRL {Phys. Rev. Lett.}
\def\PhRp {Phys. Rep.}
\def\PTP {Prog. Theor. Phys.}
\def\PTPS {Prog. Theor. Phys. (Suppl.)}
\def\NCim {Nuovo Cimento}
\def\NuPB {Nucl. Phys.}
\def\TMF {Teor. Mat. Fiz.}
\def\TMP {Theor. Math. Phys.}
\def\prp {report}
\def\Prp {Report}



\def\jn#1#2#3#4#5{#5 {\it #1}{#2} {\bf #3} {#4}}   
\def\boo#1#2#3#4#5{#4 {\it #1} ({#3}: {#2}){#5}}   
\def\prpr#1#2#3#4#5{{#5}{ #3}{#4}}                 

\end{document}